\documentclass[aps,prl,twocolumn]{revtex4}
\usepackage{graphicx}
\usepackage{bm}
\begin{document}
\preprint{}
\title{Precision quantum metrology and nonclassicality in
linear and nonlinear detection schemes}
\author{\'{A}ngel Rivas$^{1}$ and Alfredo Luis$^{2}$}
\email{alluis@fis.ucm.es}
\affiliation{$^1$Institut f\"{u}r Theoretische Physik,
Universit\"{a}t Ulm, Ulm D-89069, Germany.\\
$^2$ Departamento de \'{O}ptica, Facultad de Ciencias
F\'{\i}sicas, Universidad Complutense, 28040 Madrid, Spain}
\date{\today}

\begin{abstract}
We examine whether metrological resolution beyond coherent states
is a nonclassical effect. We show that this is true for linear
detection schemes but false for nonlinear schemes, and propose a
very simple experimental setup to test it. We find a
nonclassicality criterion derived from quantum Fisher information.
\end{abstract}


\maketitle

Nonclassicality is a key concept supporting the necessity of
the quantum theory. There is widespread consensus that the
coherent states $| \alpha \rangle$ are the classical side of
the borderline between the quantum and classical realms
\cite{MW,border}. In quantum metrology it is usually believed
that resolution beyond coherent states is a quantum effect,
since this is achieved by famous nonclassical probe states,
such as squeezed, number, or coherent superpositions of
distinguishable states \cite{CA}. However, this does not mean
that every state providing larger resolution than coherent
states is nonclassical.

In this work we test this belief by examining whether
metrological resolution beyond coherent states is necessarily
a nonclassical effect or not \cite{AAS}. To this end we find
a novel nonclassicality criterion derived from quantum Fisher
information. We demonstrate that the belief is true for linear
detection schemes but false for nonlinear schemes.
Nonlinear detection is a recently introduced item in
quantum metrology that has plenty of promising possibilities and is
being thoroughly studied and implemented in different areas
such as quantum optics \cite{QO,QO2}, Bose-Einstein condensates
\cite{BE,pe}, nanomechanical resonators \cite{NR}, and atomic
magnetometry \cite{AM}.

Throughout we focus on single-mode quantum light beams with
complex amplitude operators $a$ such that $[a,a^\dagger]=1$
and $a | \alpha \rangle = \alpha | \alpha \rangle$.
Resolution provided by different probe states is compared
for the same mean number of photons $\bar{n}$ that represents
the energy resources available for the measurement. We examine
the following proposition:

\textit{Proposition:} A probe state $\rho$ providing larger
resolution than coherent states $| \alpha_\rho \rangle$ with
the same mean number of photons $\bar{n}$ is nonclassical,
where
\begin{equation}
\label{link}
\bar{n} = \langle \alpha_\rho | a^\dagger a | \alpha_\rho
\rangle = | \alpha_\rho |^2 = \textrm{tr} (\rho a^\dagger a ) .
\end{equation}
A customary signature of nonclassical behavior is the failure
of the Glauber-Sudarshan $P (\alpha)$ phase-space representation
to exhibit all the properties of a classical probability density
\cite{MW}. This occurs when $P (\alpha)$ takes negative values,
or when it becomes more singular than a delta function. To test
the proposition we must specify how resolution is assessed.

\textit{Resolution.---}
In a detection scheme the signal to be detected $\chi$ is
encoded in the input probe state $\rho$ by a transformation
$\rho \rightarrow \rho_\chi$. For definiteness, we focus on
the most common and practical case of unitary transformations
with constant generator $G$ independent of the parameter
\begin{equation}
\rho_\chi = \exp \left ( i \chi G \right ) \rho
\exp \left ( - i \chi G \right ) .
\end{equation}
The value of $\chi$ is inferred from the outcomes of
measurements performed on $\rho_\chi$. The ultimate resolution
of such inference is given by the quantum Fisher information
$F_Q (\rho_\chi)$ since the variance of any unbiased estimator
$\tilde{\chi}$ is bounded from below in the form \cite{CR,BC}
\begin{equation}
\left ( \Delta \tilde{\chi} \right )^2 \geq \frac{1}{N F_Q
(\rho_\chi)} ,
\end{equation}
where $N$ is the number of independent repetitions of the
measurement.

Better resolution is equivalent to larger quantum Fisher
information, which can be expressed as \cite{BC,FI}
\begin{equation}
\label{Fi}
F_Q (\rho_\chi) = 2 \sum_{j,k}
\frac{(r_j - r_k )^2}{r_j + r_k} \left | \langle r_j |
G | r_k \rangle \right |^2 ,
\end{equation}
where $| r_j \rangle $ are the eigenvectors of $\rho$
with eigenvalues $r_j$ and the sum includes all the cases
with $r_j + r_k \neq 0$. So for uniparametric unitary
transformations $F_Q$ is independent of $\chi$ \cite{FI}.

In order to reach ultimate sensitivity predicted by the
quantum Fisher information, an optimum measurement and an
efficient estimator are required \cite{BC}. If we consider the
maximum likelihood as estimator, the number of repetitions
required to reach the efficient regime may depend on the probe
state \cite{BLC}. In order to focus on the intrinsic capabilities
of different schemes we will assume that $N$ is large enough so
that optimum conditions are reached for all cases, so that
schemes are compared by comparing their quantum Fisher
information. Note also that resolution depends also on the
duration of the measurement. Because of this any meaningful
comparison between different schemes should be done on equal-time basis.

Let us show three useful properties of the quantum Fisher
information:

i) For pure states, such as coherent states $| \alpha \rangle$,
the quantum Fisher information becomes proportional to the variance
of the generator \cite{BC}
\begin{equation}
\label{var}
F_Q (| \alpha \rangle ,G) = 4 \left ( \Delta_\alpha G
\right )^2 = 4 \left ( \langle \alpha | G^2 | \alpha \rangle -
\langle \alpha | G | \alpha \rangle^2 \right ).
\end{equation}

ii) The quantum Fisher information is convex. For a
proof based on the monotonocity of quantum Fisher information
under complete positive maps see Ref. \cite{FU}.  A much
simpler proof is given by a straightforward use of the
convexity of the Fisher information and the
Braunstein-Caves inequality \cite{BC}. Thus, for classical
states
\begin{equation}
\rho_\mathrm{class} = \int d^2 \alpha P_\mathrm{class} (\alpha)
| \alpha \rangle \langle \alpha |,
\end{equation}
where $ P_\mathrm{class} (\alpha)$ is a non-negative function no
more singular than a delta function, convexity implies the
following bound for the quantum Fisher information of classical
states
\begin{eqnarray}
F_Q (\rho_\mathrm{class} ,G) &\leq& \int d^2 \alpha
P_\mathrm{class} (\alpha) F_Q (| \alpha \rangle ,G) \nonumber\\
&=& 4\int d^2 \alpha P_\mathrm{class} (\alpha) \left ( \Delta_\alpha
G \right )^2.\label{conv}
\end{eqnarray}

iii) In most cases it is rather difficult to compute analytically
$F_Q (\rho ,G)$, especially in infinite dimensional systems.
A similar but simpler performance measure is
\begin{equation}
\label{spm}
\Lambda^2 \left ( \rho,G \right ) = \textrm{tr}
\left ( \rho^2 G^2 \right ) - \textrm{tr} \left ( \rho
G \rho G \right ) ,
\end{equation}
or, equivalently, in the same conditions of Eq. (\ref{Fi}),
\begin{equation}
\label{L}
\Lambda^2 \left ( \rho,G \right ) = \frac{1}{2} \sum_{j,k}
\left ( r_j - r_k \right )^2 \left | \langle r_k | G | r_j
\rangle \right |^2 ,
\end{equation}
which for pure states such as coherent states also becomes
the variance of the generator $\Lambda ( | \alpha \rangle,
G )= \Delta_\alpha G$ \cite{RL}. This is derived from
the Hilbert-Schmidt distance between $\rho_\chi$ and $\rho$
in the same terms in which the quantum Fisher information is
derived from the Bures distance \cite{BC,BF}. The useful
point here is that from Eqs. (\ref{Fi}) and (\ref{L}) and
given that $r_k + r_\ell \leq 1$ it holds that
\begin{equation}
\label{FL}
F_Q \left ( \rho,G \right ) \geq 4 \Lambda^2 \left ( \rho,G
\right ) ,
\end{equation}
the equality being reached for pure states.

\textit{Nonclassicality from quantum Fisher information.---}
For the sake of convenience let us express the variance of
$G$ on coherent states as a mean value
\begin{equation}
\label{DGAG}
\left ( \Delta_\alpha G \right )^2 = \langle \alpha | A_G |
\alpha \rangle , \quad
A_G = G^2 - :G^2: ,
\end{equation}
where $: \; :$ denotes normal order, and $G$ in $:G^2:$ must
be expressed in its normally ordered form so that $\langle
\alpha |:G^2: | \alpha \rangle = \langle \alpha |G | \alpha
\rangle^2$ . A key point is that $ \langle \alpha | A_G |
\alpha \rangle$ gives the quantum Fisher information of coherent
states,
\begin{equation}
\label{FAc}
F_Q (| \alpha \rangle ,G) = 4 \langle \alpha | A_G |
\alpha \rangle ,
\end{equation}
so that the bound (\ref{conv}) for the quantum Fisher information
of classical states reads
\begin{eqnarray}
F_Q (\rho_\mathrm{class} ,G) &\leq& 4 \int d^2 \alpha
P_\mathrm{class} (\alpha)  \langle \alpha | A_G | \alpha
\rangle  \nonumber\\
&=& 4 \textrm{tr} \left ( \rho_\mathrm{class} A_G
\right ).
\end{eqnarray}
This relation is derived from the convexity of  $F_Q (\rho ,G)$,
so it relies entirely on the classical nature of $P_\mathrm{class}
(\alpha)$. Therefore its violation provides the following
nonclassicality criterion:
\begin{equation}
\label{ncc}
F_Q (\rho ,G) > 4 \textrm{tr} \left ( \rho A_G \right )
\longrightarrow \rho \textrm{ is nonclassical}  .
\end{equation}

Since this criterion is formulated in terms of the quantum Fisher
information, it will be useful to discuss the interplay
between improved metrological resolution and nonclassicality.
The key point is to link $\textrm{tr} ( \rho A_G )$ in the
nonclassical criterion (\ref{ncc}) with the quantum Fisher
information of coherent states with the same mean number of
photons  $F_Q (| \alpha_\rho \rangle,G) = 4 \langle \alpha_\rho
| A_G | \alpha_\rho \rangle$. This is straightforward when
$A_G \propto a^\dagger a$. To study this in detail let us split
the analysis in linear and nonlinear schemes.

\textit{Linear schemes.---}
By linear schemes we mean that the signal is encoded via
input-output transformations where the output complex
amplitudes are linear functions of the input ones and their
conjugates. Their generators are polynomials of $a, a^\dagger$
up to second order, embracing all traditional interferometric
techniques exemplified by the phase shifts generated by the
photon-number operator
\begin{equation}
\label{GAa}
G = A_G = a^\dagger a ,
\end{equation}
so that $G$ and $A_G$ coincide. In this case the resolution
(quantum Fisher information) provided by coherent probe states
is given by its mean number of photons
\begin{equation}
F_Q (| \alpha_\rho \rangle ,a^\dagger a ) =  4\langle
\alpha_\rho | a^\dagger a  | \alpha_\rho \rangle = 4 \left |
\alpha_\rho \right |^2 =  4 \textrm{tr} \left ( \rho  a^\dagger
a \right ) ,
\end{equation}
where we have used Eqs. (\ref{link}), (\ref{FAc}), and
(\ref{GAa}). The probe states $\rho$ providing larger resolution
than coherent states $| \alpha_\rho \rangle$ with the same mean
number of photons satisfy
\begin{equation}
F_Q (\rho ,a^\dagger a ) > F_Q (| \alpha_\rho \rangle ,a^\dagger a )
=  4 \textrm{tr} \left ( \rho  a^\dagger a \right ) ,
\end{equation}
so that from the nonclassical criterion (\ref{ncc}) they are
necessarily nonclassical states and the proposition being tested
is true.

This result also holds for other generators of linear
transformations such as $G = a \exp (i \theta) + a^\dagger
\exp (-i \theta)$, which generates displacements of the
quadratures, and $G = a^2 \exp (i \theta) +
a^{\dagger 2} \exp (-i \theta)$, which generates quadrature
squeezing, where $\theta$ is an arbitrary phase \cite{ds}.
This is because $A_G = 1$ and $A_G = 4 a^\dagger a + 2$,
respectively, so that $4 \textrm{tr} ( \rho A_G ) = F_Q (
| \alpha_\rho \rangle , G)$.

This also holds for two-mode SU(2) generators
\begin{equation}
\label{GJ}
G = \bm{u} \cdot \bm{J} , \quad A_G = a^\dagger_1 a_1 +
a_2^\dagger a_2 ,
\end{equation}
where $\bm{u}$ is a three-dimensional unit real vector and
$\bm{J}$ are the bosonic realization of the angular momentum
operators that generate the SU(2) group
\begin{equation}
J_x = a^\dagger_1 a_2 +  a_1 a_2^\dagger,
J_y = i ( a^\dagger_1 a_2 - a_1 a_2^\dagger ),
J_z = a^\dagger_1 a_1 - a_2^\dagger a_2 .
\end{equation}
This describes all two-beam lossless optical devices, such as
beam splitters, phase plates, and two-beam interferometers.
In this two-mode context the coherent states $| \alpha
\rangle$ refer to the product of single-mode coherent states
$| \alpha \rangle = | \alpha_1 \rangle | \alpha_2 \rangle$ with
mean number of photons  $\bar{n}=|\alpha_1|^2+|\alpha_2|^2 =
\langle \alpha | A_G | \alpha \rangle$.
For a simple derivation of $A_G$ in Eq. (\ref{GJ}), note that
any $\bm{u} \cdot \bm{J}$ is in normal order, normal order
commutes with SU(2) transformations, $\bm{u} \cdot \bm{J}$
is SU(2) equivalent to $J_z$, with $J_z^2 -: J_z^2 : =
a^\dagger_1 a_1 + a_2^\dagger a_2$, and $a^\dagger_1 a_1
+ a_2^\dagger a_2$ is SU(2) invariant. This is $A_{UGU^\dagger}
= U A_G U^\dagger$ if $G$ is in normal order and $U$ is a
SU(2) unitary transformation.

When the angular momentum $\bm{J}$ refers collectively
to a system of qubits, it has been demonstrated \cite{PS} that
improved resolution beyond coherent states implies entanglement
between qubits. We recover this result by noticing that spin
nonclassicality is equivalent to entanglement \cite{eqb}.
This equivalence no longer holds when entanglement refers to the
entanglement between field modes; this is to say that nonclassical
factorized states $| \psi_1 \rangle | \psi_2 \rangle$, where $|
\psi_j \rangle$ is in mode $a_j$, can provide better resolution
than coherent states.

\textit{Nonlinear schemes.---}
By nonlinear detection schemes we mean that the signal is encoded
via input-output transformations where the output complex amplitudes
are not linear functions of the input ones. A suitable example is
given by
\begin{equation}
\label{GAGnl}
G = \left ( a^\dagger a \right )^2,  \qquad
A_G = 4 a^{\dagger 3} a^3 + 6 a^{\dagger 2} a^2 + a^\dagger a ,
\end{equation}
and the key point is that $A_G$ is no longer proportional to the
number operator. In practical quantum-optical terms this
corresponds to light propagation through nonlinear Kerr media
\cite{MW}.

Next we show that there are classical states that provide larger
resolution than coherent states with the same mean number of
photons, so that the proposition being tested is false. To this
end let us consider the mixed probe state
\begin{equation}
\label{rclass}
\rho_\mathrm{class} = p | \alpha /\sqrt{p} \rangle
\langle \alpha /\sqrt{p}| + (1-p)| 0 \rangle \langle 0 |,
\end{equation}
where $| \alpha /\sqrt{p} \rangle$ is a coherent state, $| 0
\rangle$ is the vacuum, and $1 > p > 0$. The state
$\rho_\mathrm{class}$ has the same mean number of photons
as the coherent state $| \alpha \rangle$ for every $p$.

Since in general $F_Q ( \rho_\mathrm{class},G )$ is difficult to
compute when $\rho_\mathrm{class}$ is mixed, we resort to
Eq. (\ref{FL}) so that if
\begin{equation}
\label{cond}
4 \Lambda^2 ( \rho_\mathrm{class} ,G) > F_Q (| \alpha \rangle ,G) ,
\end{equation}
then $F_Q (\rho_\mathrm{class} ,G) > F_Q (| \alpha \rangle ,G)$ and
$\rho_\mathrm{class}$ provides larger resolution than $| \alpha
\rangle$. Using Eq. (\ref{spm}) the condition (\ref{cond}) is
equivalent to the following relation between variances of $G$ in
coherent states
\begin{equation}
\label{apa}
p^2 \left ( \Delta_\frac{\alpha }{\sqrt{p}} G \right )^2 >
\left ( \Delta_\alpha  G \right )^2 ,
\end{equation}
where we have used that $| 0 \rangle$ is an eigenstate of $G$
with null eigenvalue. After Eqs. (\ref{DGAG}) and (\ref{GAGnl})
\begin{equation}
 \left ( \Delta_\alpha  G \right )^2 = 4 | \alpha |^6 +
6 | \alpha |^4 + | \alpha |^2 ,
\end{equation}
and from Eq. (\ref{apa}) the state $\rho_\mathrm{class}$
provides larger resolution than $| \alpha \rangle$ provided
that $|\alpha |^2 > \sqrt{p}/2$, which can be easily fulfilled.

We are able to observe this improvement even with a very simple
and practical measuring scheme such as homodyne detection illustrated
in Fig. 1. For that we evaluate the Fisher information
$F_C ( \rho_\mathrm{class},G )$ of the measurement for
$\rho_\mathrm{class}$ in Eq. (\ref{rclass}),
\begin{equation}
F_C ( \rho_\mathrm{class},G ) = \int dx
\frac{1}{P ( x| \chi )} \left ( \frac{ \partial
P ( x | \chi)}{\partial \chi} \right )^2 ,
\end{equation}
where $P ( x| \chi ) = | \langle x | \rho_\chi | x \rangle
|^2$ is the probability of the outcome $x$ of the $X$ quadrature, 
with $X = a^\dagger + a$ and $X |x \rangle = x | x \rangle $.
We consider very small $\chi$ so that the classical Fisher
information is evaluated at $\chi=0$. We also assume an optimum
value for the phase of the coherent amplitude $\alpha = i
\sqrt{\bar{n}}$. Using the results in Ref. \cite{QO2} we get
for large $\bar{n}$
\begin{equation}
F_C ( \rho_\mathrm{class},G) = 16 \frac{\bar{n}^3}{p} =
\frac{1}{p} F_C ( | \alpha \rangle ,G) .
\end{equation}
Thus, the Fisher information for the classical probe state
$\rho_\mathrm{class}$ is above the value for the coherent
states with the same mean number of photons $| \alpha \rangle$,
especially when $p \rightarrow 0$.
\begin{figure}
\begin{center}
\includegraphics[width=6cm]{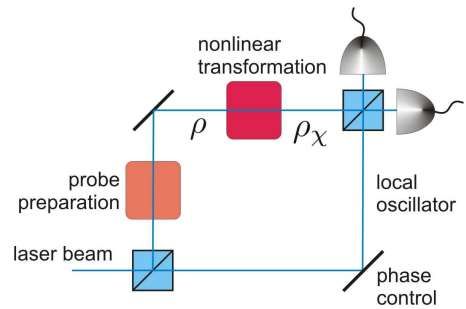}
\end{center}
\caption{Sketch of a homodyne measurement.}
\end{figure}

\textit{Discussion.---}
To some extent this may be regarded as a paradoxical result,
especially in the limit $p \rightarrow 0$ where
$\rho_\mathrm{class}$ tends to be the vacuum,
$\langle 0 | \rho_\mathrm{class} | 0 \rangle \rightarrow 1$,
since the vacuum state is useless for detection. Nevertheless
next we show that this is a fully meaningful and worthy result.
To this end let us consider that we repeat the
measurement $N$ times with the probe $\rho_\mathrm{class}$
in Eq. (\ref{rclass}). That will be equivalent to get $Np$
times the result of the probe state $| \alpha /\sqrt{p} \rangle$
and $N(1-p)$ times the useless vacuum. Therefore the useful
resources are $N | \alpha |^2$ photons distributed in $Np$ runs
of $| \alpha |^2 /p$ photons. When the probe is $| \alpha \rangle$
(this is the case $p=1$), all runs are useful and we get the same
resources $N | \alpha |^2$ distributed in $N$ runs of
$| \alpha |^2$ photons. For linear detection schemes
the two allocations of resources provide essentially the
same resolution for every $p$ because for large number of
photons $\langle \alpha | 0 \rangle \simeq 0$ it holds that $F_Q
(\rho_\mathrm{class}, a^\dagger a ) \simeq p F_Q (| \alpha
/\sqrt{p} \rangle, a^\dagger a ) = F_Q (| \alpha \rangle,
a^\dagger a )$. However, the nonlinearity greatly privileges
large photon numbers so that the best strategy is to put as
many photons as possible in a single run, instead of splitting
them into several runs. More specifically, for large $|\alpha |$
it holds that $\langle \alpha | 0 \rangle \simeq 0$ and
$F_Q (\rho_\mathrm{class} ,G) \simeq 16 |\alpha |^6 /p^2$ while
$F_Q (| \alpha \rangle  ,G) \simeq 16 |\alpha |^6 $ so that
$\rho_\mathrm{class}$ provides much larger resolution than
$| \alpha \rangle$ as $p \rightarrow 0$.

Incidentally, the above calculus shows that when $\langle
\alpha | 0 \rangle \simeq 0$ we get $F_C ( \rho_\mathrm{class}
,G ) \simeq p F_Q ( \rho_\mathrm{class},G )$. This is to say
that whereas both $F_{C,Q}$ increase when $p$ decreases, it
holds that $F_Q$ increases faster than $F_C$.

Finally, it might be argued that the improvement of resolution
in nonlinear schemes, and the differences between different classical
input probes just discussed, may be ascribed to nonclassicality
induced by nonlinear transformations. We can rule out this
possibility. The quantum Fisher information does not depend
on the value of the signal, so that the optimum sensitivity cannot
depend on the amount of nonclassicality induced by the
transformation. In particular, for the usual case of small
signals the induced nonclassicalities will be negligible.

\textit{Conclusions.---}
We have obtained a general nonclassical test derived from
quantum Fisher information. For linear detection schemes this
test demonstrates that improved resolution beyond coherent
states is a nonclassical feature. For nonlinear schemes the
situation is different since mixed classical states can provide
better resolution than coherent states.
This result is very attractive since the key point of classical
states is that they are extremely robust against experimental
imperfections \cite{QO2,pe} and they are easy to generate in labs.

\bigskip

A. R. acknowledges Susana F. Huelga for illuminating comments
and financial support from the EU Integrated Projects QAP,
QESSENCE, and the STREP action CORNER. A. L. acknowledges
support from official Spanish projects No. FIS2008-01267
and QUITEMAD S2009-ESP-1594.

\end{document}